\documentclass[10pt,a4paper]{emulateapj}
\usepackage{lmodern}
\usepackage[latin1]{inputenc}
\usepackage[english]{babel}
\usepackage{amsmath}
\usepackage{amssymb}
\usepackage{latexsym}
\usepackage{epsfig}
\usepackage{subfigure}
\usepackage{natbib}
\usepackage{setspace}
\usepackage{longtable}
\usepackage{graphicx}
\usepackage{adjustbox}
\usepackage{changepage}
\usepackage{rotating}
\usepackage{blindtext}
\usepackage{enumitem}
\usepackage{xcolor}
\usepackage[11pt]{moresize}
\usepackage{hyperref}
\hypersetup{backref=true, pagebackref=true, hyperindex=true, colorlinks=true,                
	breaklinks=true, urlcolor= magenta, linkcolor= red, bookmarks=true,                 
	bookmarksopen=false, filecolor=cyan, citecolor=blue, linkbordercolor=blue}
\usepackage{hyperref}

\newcommand{\wfirst}{\textit{Roman}}

\newcommand{\gaia}{\textit{Gaia}}

\shorttitle{Mass-velocity dispersion relation and its effect on microlensing} \shortauthors{Sajadian et al.}
\begin{document}
\title{Mass-velocity dispersion relation by using the \gaia~data and its effect on interpreting short-duration and degenerate microlensing events}

\author{Sedighe Sajadian \altaffilmark{1}, Sohrab Rahvar \altaffilmark{2}, Fatemeh Kazemian \altaffilmark{1}}
\altaffiltext{1}{Department~of~Physics,~Isfahan~University~of~Technology,~Isfahan~84156-83111,~Iran~  \email{Email: s.sajadian@iut.ac.ir} }
\altaffiltext{2}{Department~of~Physics,~Sharif~University~of~Technology,~P. O. Box~11365-9161,~Tehran,~Iran}

\begin{abstract}
Gravitational microlensing, the lensing of stars in the Milky Way galaxy with other stars, has been used for exploring compact dark matter objects, exoplanets, and black holes. The duration of microlensing events, the so-called Einstein crossing time, is a function of distance, mass, and velocities of lens objects. Lenses with different ages and masses might have various characteristic velocities inside the galaxy and this might lead to our misinterpretation of microlensing events. In this work, we use the \gaia~archived data to find a relation between the velocity dispersion and mass, and the age of stars. This mass-velocity dispersion relation confirms the known age-velocity relation for early-type and massive stars, and additionally reveals a dependence of stellar velocity dispersion on the mass for low-mass and late-type stars at $2$-$3$ sigma level. By considering this correlation, we simulate short-duration microlensing events due to brown dwarfs. From this simulation, we conclude that lens masses are underestimated by $\sim 2.5$-$5.5\%$ while modeling short-duration and degenerate microlensing events with the Bayesian analysis.  
\end{abstract}


\section{Introduction}	
The gravitational lensing happens when the light of a background source passes through the gravitational field of a foreground object which results in the bending of light \citep{Einstein1936}. On the scale of our galaxy, from gravitational bending of stellar light beams, two distorted images form where the angular separation of images is too small to be resolved by ground-based telescopes. In this case, the overall flux received by the observer is magnified. This phenomenon is known as a gravitational microlensing event \citep{Liebes1964, Chang1979, Pac1986}. As a result, a temporary and achromatic enhancement in the brightness of a background star reveals the existence of a lens passing through our line of sight to a source star  \citep{gaudi2012,rahvar2015,2018Tsapras}. This method has been used for a few decades for the detection and characterization of dark objects, even the first isolated black hole \citep{sahu}.

The timescale of a microlensing event, $t_{\rm E}$, is proportional to the square root of the lens mass. Therefore, low-mass lens objects in the Galactic disk (for example, brown dwarfs, free-floating exoplanets, wide-orbit planets, etc.) produce short-duration microlensing events \citep{Hanetal.2005,2021MNRASsa}. Detecting these short-duration microlensing events is an important tool for determining the abundance of low-mass objects in our galaxy \citep{2012NaturCassan,2018Udalski,Suzuki2016}.

The first report on the detection of short-duration microlensing events was published by {\it The Microlensing Observations in Astrophysics} (MOA) microlensing group \citep{moa2001, Sumi2003}. They discovered an excess of short-duration microlensing events \citep{Sumi2011Natur}. {\it The Optical Gravitational Lensing Experiment} group (OGLE) \citep{OGLE_IV, OGLE2003_1} announced a lower number of short-duration microlensing events in 2017 \citep{2017Mroznature}. Furthermore, several short-duration microlensing events have been discovered in recent years by OGLE and the Korea Microlensing Telescope Network surveys (KMTNet) \citep{KMTNet2016,2019Mroz_1, 2020Mroz_1, 2020Kim_1,2020Han_1, 2021Ryu_1}. In addition, \textit{The Nancy Grace Roman Space Telescope} (\wfirst)~survey is planned to search for short-duration microlensing events towards the Galactic bulge with the improved $15$-min cadence and with a high photometric precision \citep{Penny2019,Penney2020,2019Bagheri,2021sajadianHabit}. 

One of important factors in the interpretation of short-duration events is understanding the velocity of these objects in the galaxy. In this work, we investigate the values of the velocity dispersion of low-mass stars compared to other stars in the Milky Way. This would change the interpretation of short-duration microlensing events. Using the \gaia~archived data, we first search for any correlation between stellar mass and the distribution of their velocity dispersion and show that there is a correlation between the scale parameter of the velocity dispersion profile, $"a"$, and the mass of stars which is $a \propto 1/\sqrt{M}$. This means that low-mass stars have a wider velocity dispersion profile. Since there is also a correlation between the age and the mass of stars, this correlation confirms the known age-velocity dispersion relation and manifests an extra correlation between stellar velocity dispersion and mass. We evaluate the effect of this correlation on the interpretation of microlensing events.

In section \ref{three}, we first revisit the well-known age-velocity dispersion relation. Then, using the \gaia~archived data, we look for any correlation between mass and velocity dispersion of stars. In section \ref{four}, we study the effect of this correlation on the interpretation of short-duration and degenerate microlensing events. We will summarize the results in section \ref{five}.

\section{Dispersion velocity of stars}\label{three}

In our galaxy, all stars have two types of velocities: global and dispersion velocities. The global velocity results from the mean gravitational potential of the galaxy. For instance, in the spiral galaxies all stars rotate around the galactic center independent of their masses. This global velocity increases by getting away from the galactic center up to the edge of the galactic bulge and then gets flat. On the other hand, the velocity dispersion results from the gravitational interaction of stars with the galactic giant structures and neighboring stars. However, the interaction with large structures, for instance, gas clouds, spiral arms, and stellar clusters, make the main disk heating mechanism for all stars regardless of their masses \citep{BennyTbook}. For that reason, at first glance, we do not expect any dependence of stellar velocity dispersion on their masses at a given age. On the other hand, star-star scattering or star interactions with hypothetical Massive Compact Halo Objects (MACHOs) cause stellar velocity dispersion depending on their masses. In the following subsection, we revisit the known relation between velocity dispersion and stellar age.

\subsection{The known age-velocity dispersion relation}

Stars in spiral galaxies, while having orbital global velocity around the galactic centers, have three local dispersion velocities in the radial (toward to the Galactic center), azimuthal (or tangential) and vertical directions, namely $v_{\rm U}$, $v_{\rm V}$, and $v_{\rm W}$, respectively. There are several known mechanisms which are responsible for the time evolution of these dispersion velocity components, which are mostly related to irregularities in the galactic gravitational potential and its variations with time. In the following, we briefly explain some of them.

Massive gas or molecular clouds can influence the stellar orbits in the galactic disk and, as a result, encounter orbiting stars. In that case, the difference between values of stellar angular speeds and the cloud's velocity makes some stellar epicycle motions which warms the galactic disk. However, this mechanism affects radial dispersion velocities more than vertical ones \citep{1953Spitzer,1984Lacey}.

Spiral arms in our galaxy can scatter stars, because the galactic disk is not uniform and there are several concentrated features including young stars and clouds inside spiral arms. These features are described by a sinusoidal gravitational potential which imposes an epicyclic motion with radial and azimuthal oscillations on stars. If these spiral patterns change with time, they cause heating of the galactic disk \citep{BennyTbook}. However, these transient spiral patterns, unlike the galactic bar, can not efficiently change the stellar vertical velocities, because their time scale is long compared to the time scale of the stellar oscillations vertical to the galactic disk \citep{2010Saha,2016Monari}.

These two factors are known as the main mechanisms for the time evolution of the stellar dispersion velocities in the galactic disk. This time evolution is known as the Age-Velocity dispersion Relation (hereafter AVR) and it is formulated as $v_{i} \propto \mathcal{A}^{\beta}$. Here, $\mathcal{A}$ refers to the stellar age, $v_{i}$ represents different components of velocity dispersion, i.e., $i \in U,~V,~W$ , and $\beta \in [0.3,~0.5]$ \citep{1974Mayor, 1974Wielen, Carlberg1985, 1998Dehnen}. However, the relationships between stellar age and components of dispersion velocity have been confirmed in the literature using stellar kinematic data collected by surveys such as the Hipparcos, \gaia~telescopes and the Geneva-Copenhagen survey \citep{Gomez1997, Jincheng2018, Mackereth2019, Nordstrom2004}.

Although these two mentioned mechanisms have main contributions to the time evolution of velocity dispersion in the galactic disk, there are some other second-order irregularity sources in the disk gravitational potential. For instance, either MACHOs (e.g., black holes) or other field stars can potentially encounter disk stars. Although these encounters happen over the time scale as long as the galactic age, they potentially cause either disk heating or cooling depending on their velocities. MACHO interactions evolve radial and vertical velocities of dispersion similarly with time as $\mathcal{A}^{0.5}$ \citep{BennyTbook}. Since real observations do not confirm similar time-dependence for radial and vertical components of stellar dispersion velocities, MACHO interactions can not be the main source for the AVR.

The AVR is also included in the dynamical models for the galaxy. For instance, in the Besan\c{c}on model \footnote{\url{https://model.obs-besancon.fr/}} the AVR was considered for different components of velocity dispersion \citep{Robin2003, Robin2012}. In this model, by increasing stellar age from $1$ to $10$ Gyr, the velocity dispersion changes from $\sim30$ to $\sim60~\rm{km/s}$ \citep{Bienayme1987, Robin2003}.

Since massive stars have shorter lifetimes and low-mass stars have longer lifetimes, we expect to have a correlation between the mass and the age-velocity relation. The stellar lifetime, $\tau$, depends on the mass of stars as \citep{Maeder1989,Tinsley1980,Tosi1982}:
\begin{eqnarray}
\tau \simeq \tau_{\odot} (\frac{M}{M_{\odot}})^{-2.5}, 
\end{eqnarray}
\noindent where, $\tau_{\odot}$ is the lifetime of the Sun. Hence, an output of the known AVR is that early-type stars on average have smaller dispersions than late-type stars. 

According to the known AVR, the components of the dispersion velocity increases as a power-law function by the age of star (i.e. $v_{i} \propto \mathcal{A}^{\beta}$) where exponent ranges $\beta \in[0.3,~0.5]$ and the rate of decreasing the dispersion velocity with the age as $dv_{i}/d\mathcal{A} \propto\mathcal{A}^{\beta-1}$ with $\beta-1<0$ \citep{Carlberg1985}. We note that for late-type stars with masses of less than one solar mass, e.g., K- and M-type stars, and brown dwarfs, their lifetimes are longer than the age of the Universe. Hence, their average ages are similar and the AVR predicts similar dispersion profiles for them. In the next part, we will use the \gaia~data to test this hypothesis.

\subsection{The mass-velocity dispersion relation}

As we discussed in the previous section, the gravitational interactions between stars in the galaxy and its giant structures, such as gas and molecular clouds and transient spiral arms, are main mechanisms for the time evolution of velocity dispersion components, i.e., AVR. This AVR does not depend on the stellar mass, because gravitational potential due to these giant structures makes the same acceleration for different masses and causes an overall disk heating for all stars. Other irregularity sources in the disk potential have smaller contributions in the AVR.\\

One of these sources is the encountering of disk stars with MACHO objects or each other. We note that this kind of interaction causes a mass-dependent evolution for velocity dispersion. Since the time scale of this mechanism is too long and in the other of the galactic age, theoretical studies predict a negligible contribution due to this heating mechanism in the AVR. Nevertheless, in this work we aim to examine this point and evaluate this contribution by probing any potential mass-dependence of stellar velocity dispersions through the \gaia~archived data.\\

The distribution function of the stars inside a collisional system can be given with the Maxwell-Boltzmann (MB) distribution \citep{Peckham1992, Andrew2001}, 
\begin{eqnarray}\label{MBD}
f(v) =\sqrt{\frac{2}{\pi}}~\frac{v^{2}}{a^{3}}~\exp\left(\frac{-v^2}{2 a^{2}} \right) , 
\end{eqnarray}

\noindent where, $v$ is the size of total velocity and $a$ is the scale parameter and is given as $a=\sqrt{k~T/M}$, $M$ is the mass of particles in the system and $kT$ is a constant for that system. Globular clusters in a galaxy are good examples of such thermalized systems. In these collisional systems the velocity profile of stars depends on their mass through the scale parameter. The mean and most probable velocities, $v_{\rm{m}}$, $v_{\rm{p}}$, and root-mean-square velocity, $v_{\rm{rms}}$, are all proportional to this scale parameter. The larger the scale parameter, the wider the distribution function. As a result, for a given temperature, particles with lower masses have a wider and more extended distribution of velocities, implying that low-mass stars can have higher velocities.

\noindent However, in our galaxy, encountering of stars with MACHOs or each other is rare, which causes a weak dependence stellar velocities on their masses. Here, we aim to examine such dependence (even a weak relation) through the \gaia~archived data.

\begin{figure}[pb]
\includegraphics[width=0.5\textwidth]{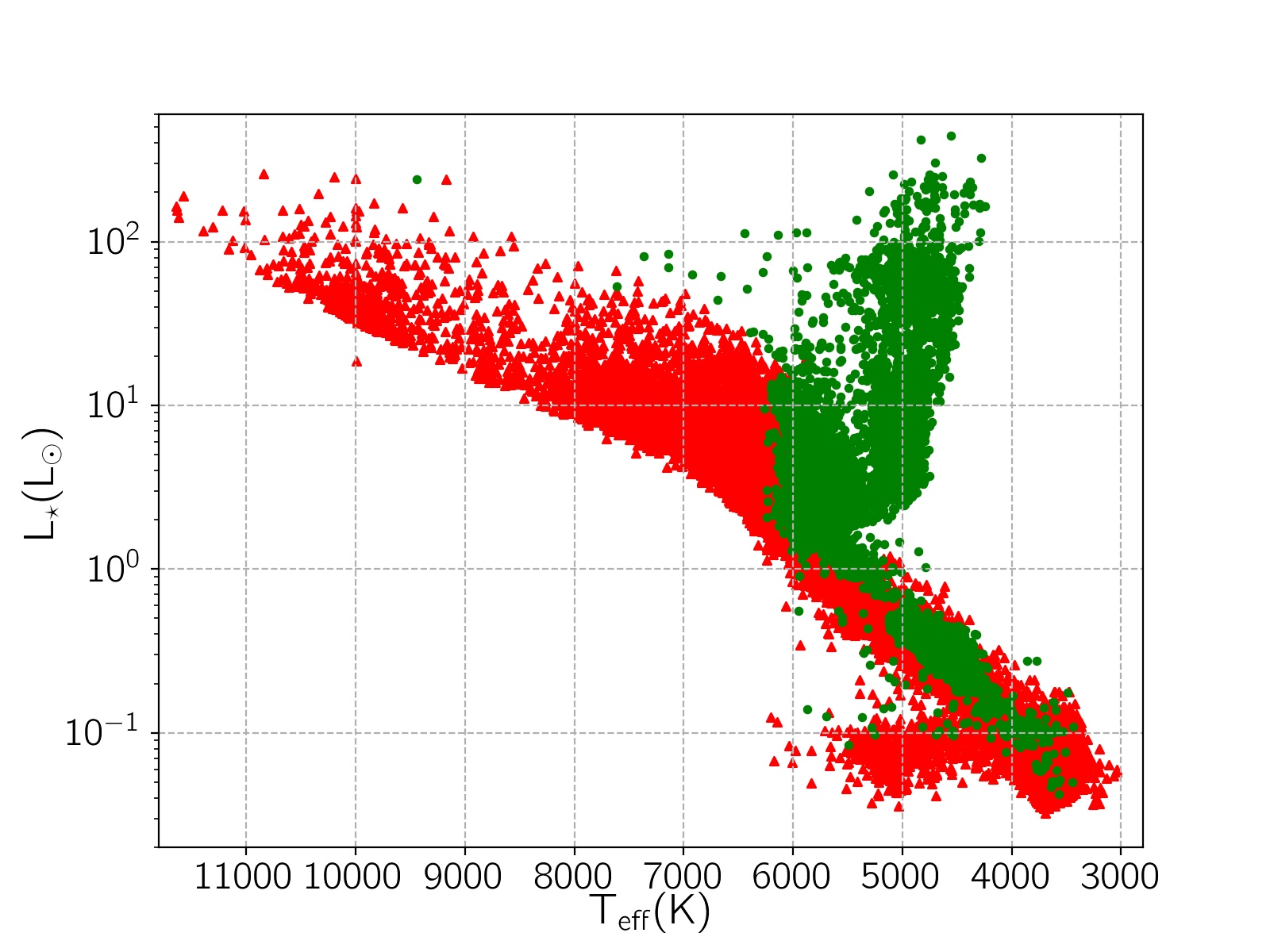}
\vspace*{8pt}
\caption{The Luminosity-Temperature diagram of stars in our sample, as reported in the \gaia~DR3. Main-sequence and giant stars are specified by red triangles and green circles, respectively.}
\label{cmd}
\end{figure}

We use the \gaia~data to obtain the distribution of stars in the Milky Way galaxy. We take a large ensemble from the \textit{Gaia} archive \footnote{\url{https://gea.esac.esa.int/archive/}} (\textit{Gaia} Data Release 3 (DR3)) \citep{2016GaiaCollaborationI,2021gaias,2022babgaia}. These stars are (a) closer than $150$ pc, (b) brighter than $21$ mag in the $G$-band (as these stars make a complete sample), and (c) additionally, their temperature, luminosity, mass, age, and their velocity components in the heliocentric frame have been measured. Also, we exclude variable or giant stars and take only main-sequence ones, to exclude very old stars. The scatter plot of the \gaia~stars in our sample over the Luminosity-Temperature diagram is shown in Figure \ref{cmd}, where main-sequence and giant stars are specified by red triangles and green circles, respectively.

\noindent These stars have similar global velocities in the neighborhood of the Sun ($\leq 150$ pc) and any peculiar velocity in the \gaia~data can be regarded as velocity dispersion of stars in the Galaxy. For all of these stars, we have the projected components of their velocities on the sky plane, towards the increase of right ascension $\alpha$ and declination $\delta$, i.e., $(v_{\alpha},~v_{\delta})$. The third component, i.e., the stellar velocity in the line of sight direction $v_{\rm r}$, has been measured with spectroscopy and reported in the \gaia~data. \\

We first transform the stellar velocity components reported by the \gaia~telescope from the heliocentric frame to the local Galactic frame, i.e., the Local Standard of Rest (LSR). For this transformation, we need the Sun's peculiar velocity components, i.e., $(v_{\rm U},~v_{\rm V},~v_{\rm W})_{\odot}=(11.1^{+0.69}_{-0.75}, ~12.24^{+0.47}_{-0.47}, ~7.25^{+0.37}_{-0.36})$~$\rm{km/s}$ \citep{2010Ralph, 2019Ding}. Therefore, we first transform the stellar velocity components reported in the \gaia~archive from the heliocentric frame $(v_{\alpha},~v_{\delta},~v_{\rm r})$ to the Galactic frame $(v_{\rm U},~v_{\rm V},~v_{\rm W})$ using the stars' celestial coordinates, i.e., $(\alpha,~\delta)$ \citep[see, e.g., ][]{Zbinden_2019}. Then, we add the Sun's peculiar velocity to these stellar velocity components, as explained in \citet{2010ApJBond, 2012Ralph,2019Everall}. The resulting velocities will be in the LSR frame. Since we need only the size of velocity dispersions, we do not convert them back to the heliocentric frame.

\begin{figure*}
\begin{center}
\includegraphics[width=0.49\textwidth]{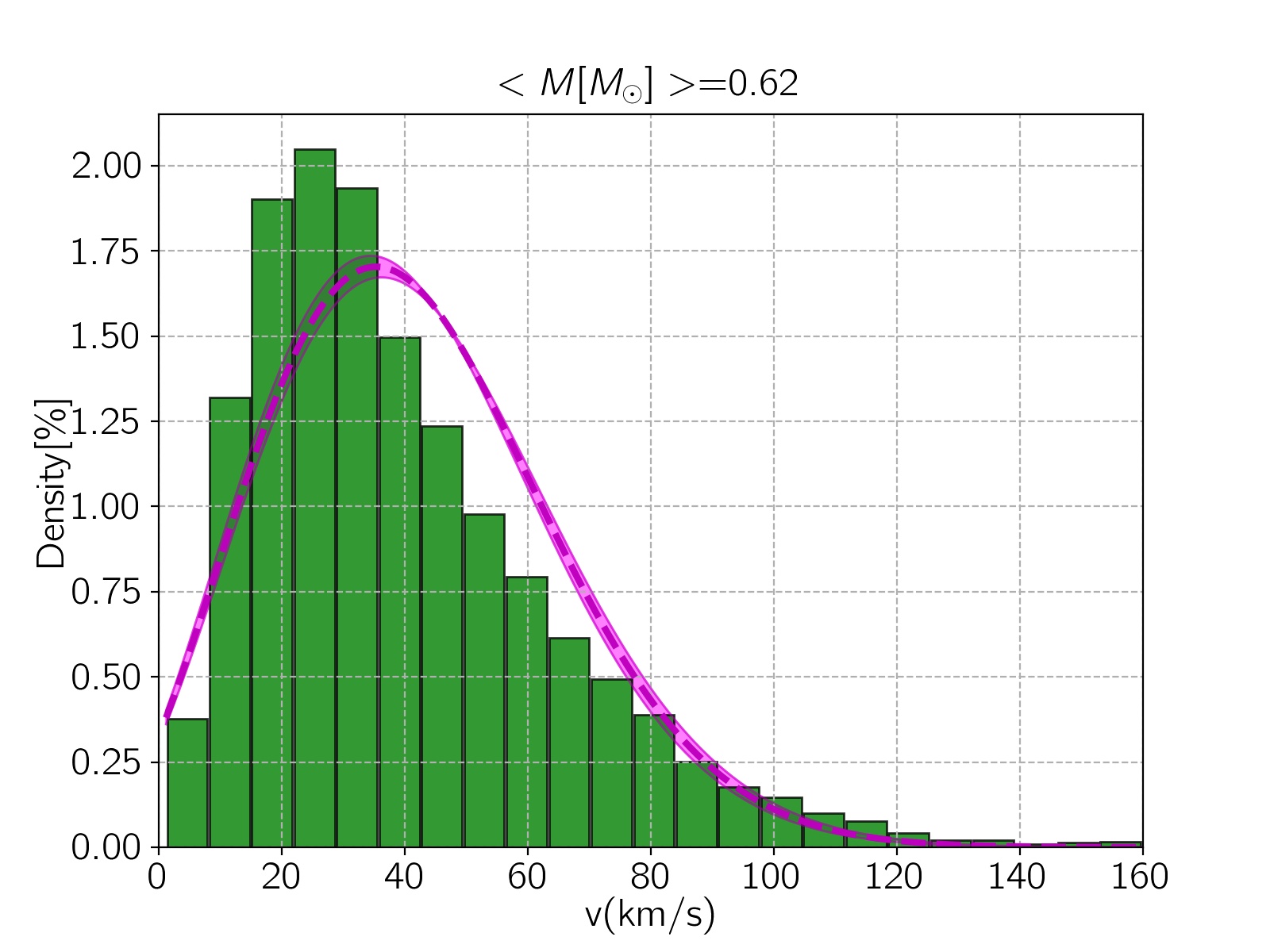}
\includegraphics[width=0.49\textwidth]{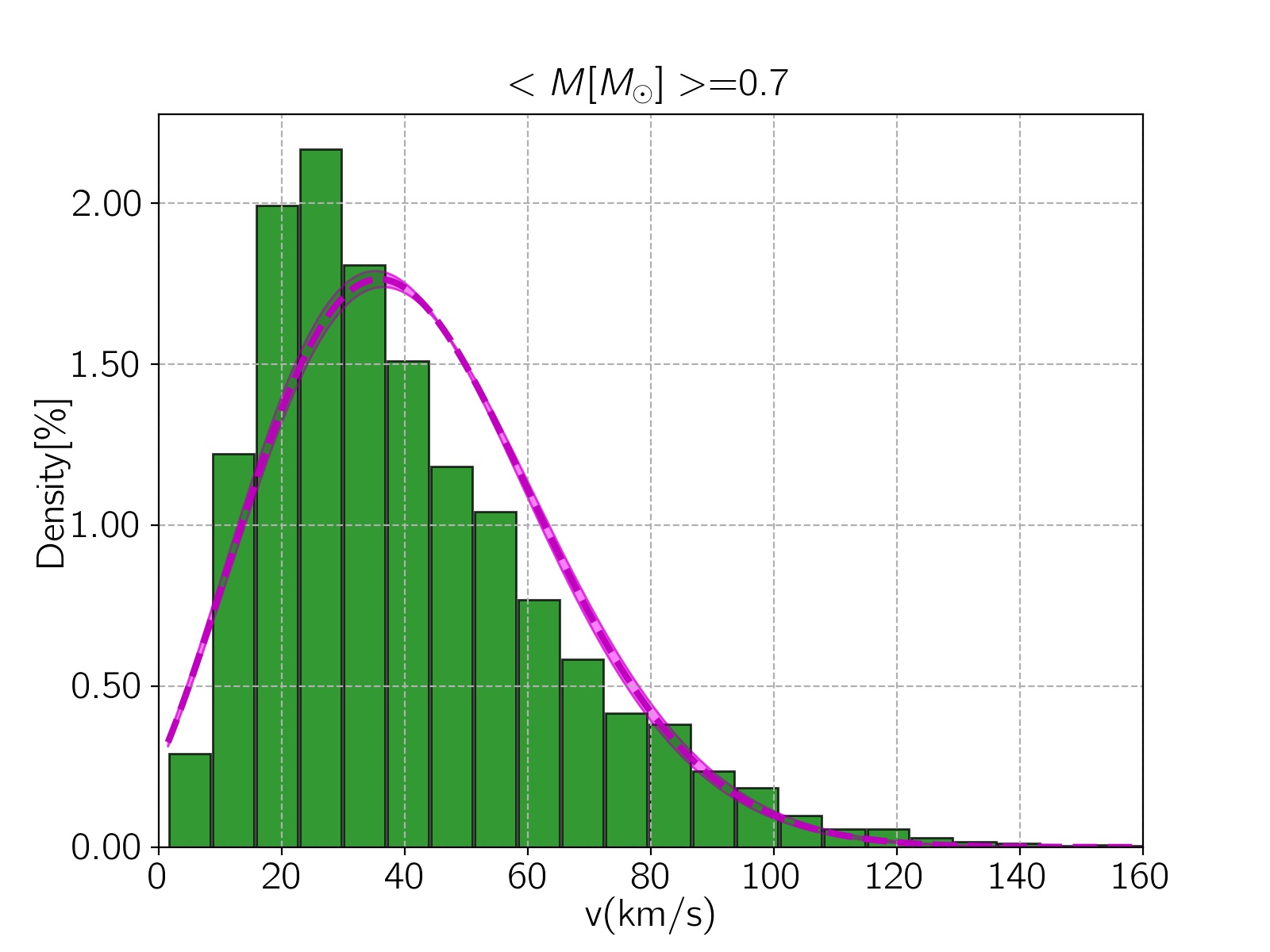}
\includegraphics[width=0.49\textwidth]{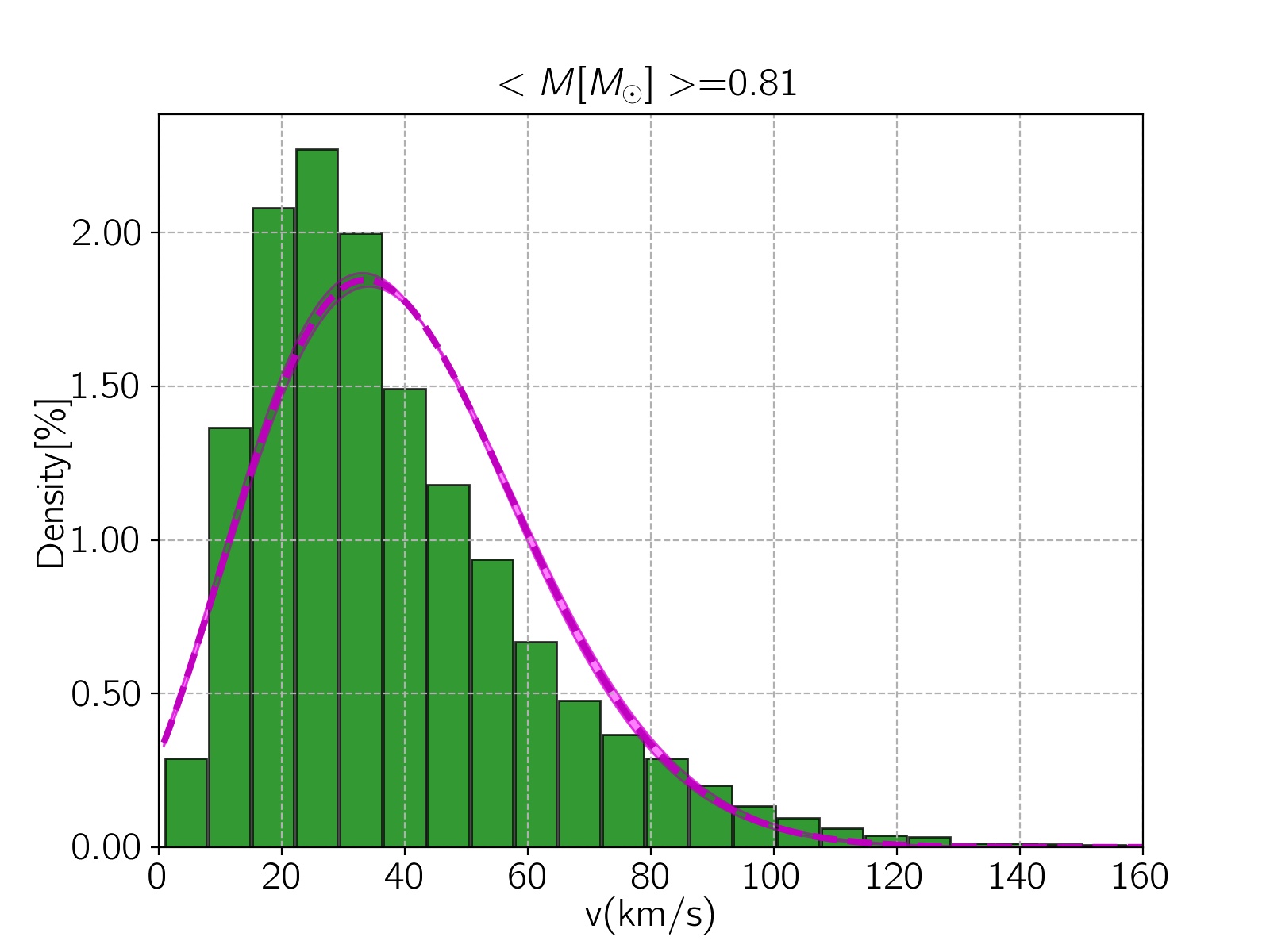}
\includegraphics[width=0.49\textwidth]{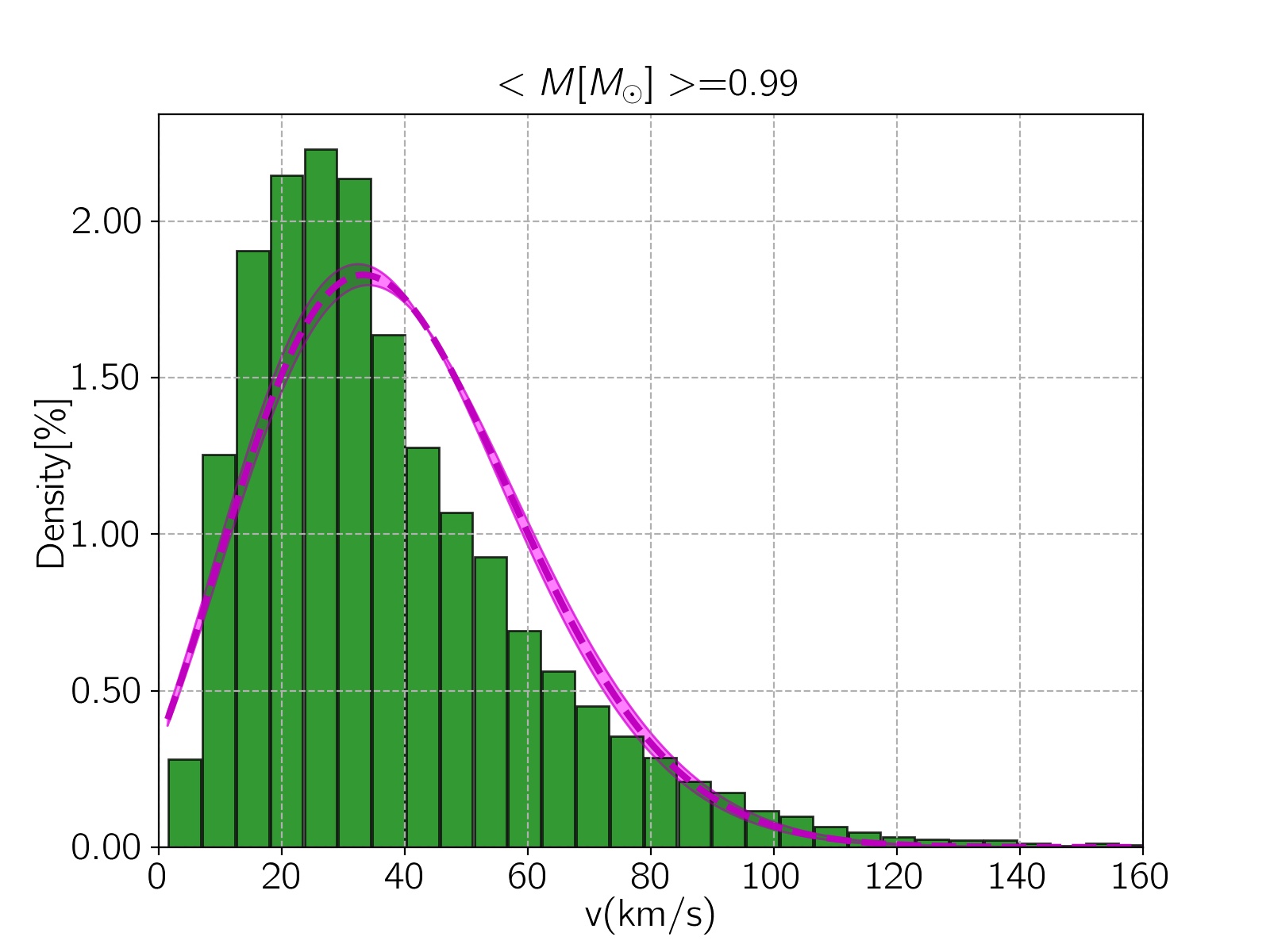}
\caption{The velocity profiles for different stellar masses detected by the \gaia~telescope. For each profile, an MB distribution is fitted to data which is shown by a magenta dashed curve. Two curves which restrict the filled magenta area represent MB distributions with the scale parameters $a \pm \sigma_{a}$. The parameters of the best-fitted MB curves are reported in Table \ref{table1}.}\label{veloc}
\end{center}
\end{figure*}

We divide our sample of stars into 10 subsamples. The mass ranges of these subsamples are~$[0.52,~0.59] M_{\odot}$, $[0.59,~0.66] M_{\odot}$, $[0.66,~0.72] M_{\odot}$, $[0.72,~0.95] M_{\odot}$, $[0.95,~1.05] M_{\odot}$, $[1.05,~1.20] M_{\odot}$, $[1.20,~1.33] M_{\odot}$, $[1.33,~1.51] M_{\odot}$, $[1.51,~1.63] M_{\odot}$, and $[1.63,~3.2] M_{\odot}$. The massive stars with $M>1.6 M_{\odot}$ are very scarce in our sample. We had to consider a larger range of mass for the massive stars to have considerable statistics in its subsample. The total number of stars in all subsamples is $\simeq 117542$.

\noindent By knowing the stellar mass and velocity components of stars in each subsample, we plot the profile of the dispersion velocity of stars. In Figure \ref{veloc}, we show four distributions of the stellar velocity dispersion in various mass ranges. These velocities are the size of dispersion velocity in the LSR frame, i.e., $v=\sqrt{v_{\rm U}^{2} + v_{\rm V}^{2} + v_{\rm W}^{2}}$. The average mass of stars for each profile is mentioned at the top of the plot. The best-fitted MB distributions are shown with magenta dashed curves. Two other curves represent the MB distributions with the scale parameters $a \pm \sigma_{a}$ \footnote{The process of fitting MB distributions to histograms is performed by the Python module \texttt{scipy.stats} with the address \url{https://docs.scipy.org/doc/scipy/reference/stats.html}.}.\\

In Table \ref{table1}, we represent the parameters of the best-fitted MB distributions. In this table, $\left< \mathcal{A}\right>$ is the average value of stellar ages in each subsample, $[\rm M/ \rm H]$ represents the stellar metallicity, $\rm{STD}$ is the square root of the variance (Standards Deviation) of distributions, and $\rm{No.}$ is the number of stars in each subsample. 

\noindent We adapt the errors in velocity values from the \gaia~database, and their average values, $\left<\sigma_{\rm v}\right>$, in different subsamples are mentioned in the fourth column of Table \ref{table1}. Using these error values, we determine the errors in the best-fitted scale parameters (in the fifth column of Table \ref{table1}). In this regard, we take many synthetic samples of stellar velocities by considering their errors. We then fit MB distributions to these samples and get a distribution for the best-fitted scale parameters. The width of this distribution determines the error of the scale parameter, $\sigma_{\rm a}$. These error bars depend on the error in velocities as measured by \gaia~and the number of stars in each subsample. The errors in velocity values mentioned by \gaia~strongly depend on the apparent stellar brightness.

\begin{deluxetable*}{c c c c c c c c c c }
\tablecolumns{10}\centering\tablewidth{0.95\textwidth}\tabletypesize\footnotesize\tablecaption{The parameters of the best-fitted MB distributions for different subsamples of stars observed by \gaia.}
\tablehead{\colhead{$\left<M\right>$}&\colhead{$\left<\mathcal{A}\right>$}&\colhead{$\left<[\rm M/ \rm H]\right>$}&\colhead{$\left<\sigma_{\rm v}\right>$}&\colhead{$\rm{a}$}&\colhead{$v_{\rm{m}}$}&\colhead{$v_{\rm{p}}$}&\colhead{$v_{\rm{rms}}$}&\colhead{$\rm{STD}$}&\colhead{$\rm{No.}$}\\
$\rm{(M_{\odot})}$&$\rm{(Gyr)}$&$\rm{(dex)}$&$\rm{(km/s)}$&$\rm{(km/s)}$&$\rm{(km/s)}$&$\rm{(km/s)}$&$\rm{(km/s)}$&$\rm{(km/s)}$&}\\
\startdata
$1.92 \pm 0.31$ &  $0.94$ & $-0.44$ & $0.93$ & $17.04 \pm 1.14$ & $27.20 \pm 1.82$ & $24.10 \pm 1.61$ & $29.52 \pm 1.97$ & $11.48$ & $5113$\\
$1.57 \pm 0.03$ &  $1.49$ & $-0.37$ & $0.75$ & $20.41 \pm 0.74$ & $32.57 \pm 1.18$ & $28.87 \pm 1.05$ & $35.36 \pm 1.28$ & $13.75$ & $1619$\\
$1.40 \pm 0.05$ &  $2.16$ & $-0.35$ & $0.65$ & $21.60 \pm 0.87$ & $34.47 \pm 1.39$ & $30.55 \pm 1.23$ & $37.41 \pm 1.51$ & $14.55$ & $4894$\\
$1.26 \pm 0.04$ &  $2.86$ & $-0.32$ & $0.61$ & $28.01 \pm 1.13$ & $44.69 \pm 1.80$ & $39.61 \pm 1.60$ & $48.51 \pm 1.96$ & $18.86$ & $5317$\\
$1.12 \pm 0.04$ &  $3.74$ & $-0.24$ & $0.50$ & $30.15 \pm 0.68$ & $48.11 \pm 1.09$ & $42.64 \pm 0.96$ & $52.22 \pm 1.18$ & $20.30$ & $10412$\\
$0.99 \pm 0.03$ &  $4.42$ & $-0.12$ & $0.47$ & $32.12 \pm 0.59$ & $51.25 \pm 0.94$ & $45.42 \pm 0.83$ & $55.63 \pm 1.02$ & $21.63$ & $15789$\\
$0.81 \pm 0.07$ &  $7.10$ & $-0.00$ & $0.62$ & $31.80 \pm 0.38$ & $50.74 \pm 0.61$ & $44.97 \pm 0.54$ & $55.07 \pm 0.66$ & $21.41$ & $35199$\\
$0.70 \pm 0.02$ &  $6.70$ & $-0.10$ & $1.24$ & $33.27 \pm 0.47$ & $53.09 \pm 0.75$ & $47.05 \pm 0.66$ & $57.62 \pm 0.81$ & $22.40$ & $11423$\\
$0.62 \pm 0.02$ &  $8.74$ & $-0.28$ & $2.35$ & $34.46 \pm 0.64$ & $54.99 \pm 1.02$ & $48.73 \pm 0.91$ & $59.69 \pm 1.11$ & $23.21$ & $11238$\\
$0.55 \pm 0.02$ &  $10.66$ & $-0.34$ & $3.18$ & $35.38 \pm 1.52$ & $56.46 \pm 2.43$ & $50.03 \pm 2.15$ & $61.28 \pm 2.63$ & $23.83$ & $16542$\\
\\
\multicolumn{10}{c}{$\rm{Extrapolated}~\rm{values}~\rm{for}~\rm{BDs}$}\\
\\
$0.04$ & $--$ & $--$ & $--$ &  $53.76\pm 12.12$ & $85.79\pm 19.34$ & $76.03 \pm 17.14$ & $93.12 \pm 20.99$ & $36.21$ & $--$\\
\\
\enddata
\tablecomments{Here, $\mathcal{A}$ refers to the stellar age, $\rm{STD}$ is the standard deviation (the square root of the variance) of each distribution, and $\rm{No.}$ is the number of stars in each subsample.}
\label{table1}
\end{deluxetable*}
\begin{figure}
\centering
\includegraphics[width=0.49\textwidth]{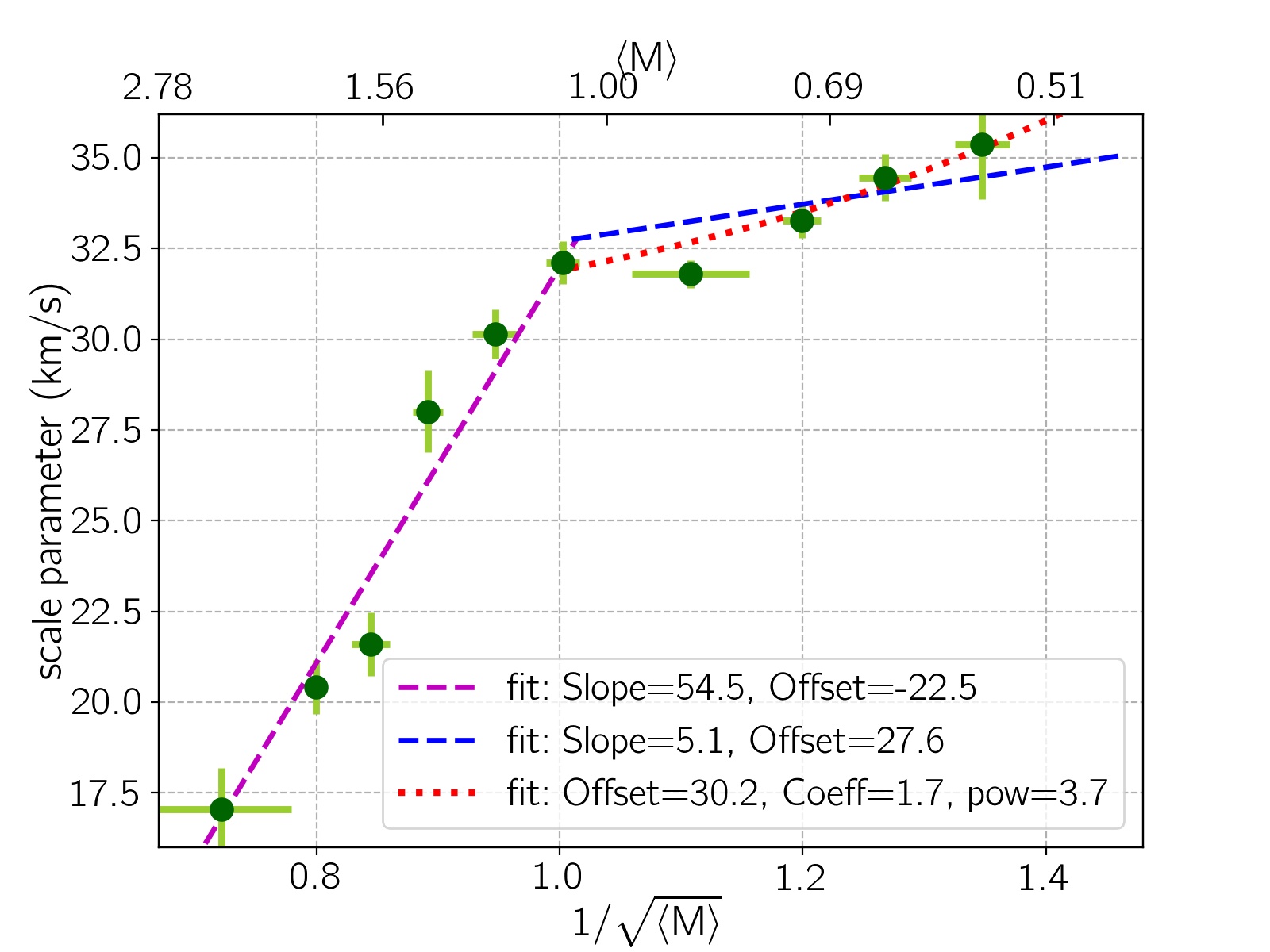}
\caption{The scale parameter versus $1/\sqrt{\left<M\right>}$ (in the unit of the solar mass) for different subsamples with different mass ranges (green points). The best-fitted linear and power-law lines, given by Equations \ref{corre} and \ref{corre2}, are shown by dashed and dotted lines, respectively.}\label{fitted}
\end{figure}

In Figure \ref{fitted}, we show the scatter plot of the scale parameters versus the inverse of the square root of the average mass of stars in different subsamples. The trend of this curve is rising. We note that the scale parameter is proportional to the inverse of the square root of the average mass of stars for a thermalized system (a linear relation). By roughly considering our galaxy as a thermalized system, we fit two linear functions for the scale parameters versus $1/\sqrt{\left<M\right>}$, which are depicted in Figure \ref{fitted} by dashed (magenta and blue) lines. The equations of these fitted lines are:
\begin{eqnarray}\label{corre}
a(\rm{km}/\rm{s})=
\begin{cases}
\frac{54.5(\pm 6.3)}{\sqrt{\left<M\right>}}-22.5(\pm 5.5) &~ \left<M\right> \gtrsim  0.98 M_{\odot}, \\  
\frac{5.1(\pm 2.4)}{\sqrt{\left<M\right>}}+ 27.6(\pm  2.8) &~  \left<M\right> \lesssim 0.98 M_{\odot}. \\
\end{cases}
\end{eqnarray}
In these equations, the mass is normalized to the solar mass and the scale parameter is in the unit of $\rm{km/s}$ \footnote{The errors in slopes, offsets and power index were estimated using the covariance matrices which have been calculated by the Python module \texttt{scipy.optimize}: \url{https://docs.scipy.org/doc/scipy/reference/optimize.html}.}.\\

We note that the mean velocity in MB distributions, $v_{\rm{m}}=\sqrt{8/ \pi}~a$, is proportional to scale parameters. As a result, we call this relationship between mass and scale parameter as \textit{mass-velocity dispersion relation} (hereafter MVR).

\noindent The best-fitted lines have positive slopes. The first part of the MVR, which is for stars with $\left<M\right> \gtrsim 0.98 M_{\odot}$, in fact confirms the known AVR. Because, more massive stars are on average younger and have smaller velocity dispersions. The second part of the MVR for $\left<M\right> \lesssim 0.98 M_{\odot}$ has a small increasing slope and shows for late-type stars their dispersion somewhat decreases with mass, although these stars have similar average ages. We note that the evidence of the MVR for late-type stars is only at $2$-$3$ sigma level.\\ 

In fact, our galaxy is not completely thermalized, since its relaxation time is of the order of $10$ Gyrs. Hence, we fit a power-law relation to the data due to late-type stars as well. According to the behavior of data in this plot, the equation of the best-fitted power-law line for $\left<M\right> \lesssim 0.98 M_{\odot}$ is: 
\begin{eqnarray}\label{corre2}
a(\rm{km}/\rm{s})=30.2 (\pm 1.5) + 1.7(\pm 1.0) \Big(\frac{1}{\sqrt{\left<M\right>}} \Big)^{3.7(\pm 4.7)}.
\end{eqnarray}
This line is shown with a dotted red curve in Figure \ref{fitted}, which is close to the dashed blue line. 

We note that the correlation between stellar mass and velocity dispersion is resulted by averaging the ages of stars in each subsample. Because the number of stars in different subsamples is relatively low, we could not make other subsamples due to different ages. For a larger sample of stars, one can study the correlation between velocity dispersion, mass and age.\\

According to Equations \ref{corre}, we expect that brown dwarfs (BD) have a wider velocity dispersion profile. To estimate the scale parameter of the MB velocity distribution for BDs, we estimate their average mass. The average mass of stars in a given mass range (for any stellar type) depends on the mass density function. The mean values for stars in different subsamples are given in the first column of Table \ref{table1}. The mass function of BDs is given by $dN/dM \propto M^{-\alpha_{0}}$, with $\alpha_{0}\simeq 0.7$ \citep{SONYCII, Luhman2004}. Accordingly, the average mass of BDs (over the range of $[13,~80]M_{\rm J}$) is $\left<M_{\rm{BD}}\right>=39.8 M_{\rm J}=0.038M_{\odot}$, where $M_{\rm J}$ is the Jupiter mass.

\noindent By considering the MVR, given in Equations \ref{corre}, we infer the expected scale parameter for BDs. The extrapolated parameters are mentioned in the last row of Table \ref{table1}. Accordingly, BDs have a wider distribution for their dispersions than main-sequence stars, which means that BDs can have higher velocity dispersions. 

In the second column of Table \ref{table1}, average stellar ages for each subsample are reported. As expected, massive stars are on average younger and, as a result, have smaller velocity dispersions. That confirms the well-known AVR. Hence, while comparing early-type stars with late-type ones, the MVR is similar to the AVR.

The MVR, on the other hand, has a noticeable impact on low-mass stars and BDs and is independent of the AVR. BDs with a much lower mass (one-tenth of late-type stars' mass where we have $1/\sqrt{\left<M_{\rm{BD}}(M_{\odot})\right>}=5.1$) have a wider dispersion profile with the scale $a \sim 54~\rm{km/s}$. The MVR has an impact on the interpretation of degenerate microlensing events (especially short-duration ones due to low-mass objects). We evaluate this effect in the next section.

\section{Impact of the MVR on microlensing interpretations}\label{four}

The timescale of microlensing events is called the Einstein crossing time, $t_{\rm E}$, which is given by \citep[see, e.g., ][]{gaudi2012, 2018Tsapras}:  

\begin{eqnarray}
t_{\rm E}= \frac{1}{v_{\rm{rel}}}\sqrt{\frac{4 G M_{\rm l} D_{\rm s}}{c^{2}}~x_{\rm{rel}}~(1-x_{\rm{rel}})},
\end{eqnarray}

\noindent where $M_{\rm l}$ is the lens mass, $D_{\rm s}$ is the source distance, $D_{\rm l}$ is the lens distance, $x_{\rm{rel}}=D_{\rm l}/D_{\rm s}$ is the ratio of the lens distance to the source distance from the observer, and $c$ is the light speed. This time scale is measurable from microlensing lightcurves. We note that $t_{\rm E}$ is a degenerate function of the lens mass, the lens and source distances from the observer and the lens-source relative velocity $v_{\rm{rel}}$ \citep[e.g., ][]{DiStefano2012}. For instance, short-duration microlensing events can be generated by either low-mass lens objects (e.g., free-floating exoplanets), or very close microlenses $(x_{\rm{rel}} \ll 1)$, or the lens objects with high transverse velocities, or a combination of these three situations. However, if the finite-source and parallax effects are measured during a microlensing event, this degeneracy is resolvable (from the observation of the galactic bulge and by knowing the source distances) \citep[see, e.g.,][]{2004Yoo,2013Nataf,Shvartzvald2019}.

According to the MVR, low-mass stars can potentially have higher velocity dispersions. Hence, a population of lens objects, by considering this correlation, generates shorter microlensing events. However, this correlation has a higher impact on short-duration microlensing events. In order to show this effect, we perform a Monte-Carlo simulation of microlensing events towards the Galactic bulge. We generate source stars population using the Besan\c{c}on model \citep{Robin2003, Robin2012}. We have mentioned all details of such simulations in the previous papers \citep{Sajadian2015, sajadian2019, Moniez2017}. Briefly, the source distance is chosen from the overall mass density versus distance in a given direction as $d^{2}M/dD_{\rm s}d\Omega \propto D_{\rm s}^{2}~\rho_{\rm{tot}}(D_{\rm s}, \alpha, \delta)$. Here, $\alpha, ~\delta$ are the right ascension and declination for a given direction and $\rho_{\rm{tot}}$ is the overall mass density in our galaxy due to all structures. The photometric properties of source stars are given by the Besan\c{c}on model.

To generate the lens population, we consider only BDs as microlens objects. Their mass is chosen from the range $M_{\rm l} \in [13 M_{\rm J},~0.08 M_{\odot}]$ with the mass function of $dN/dM \propto M^{-0.7}$ \citep{SONYCII, Luhman2004}. For lens objects, we have their mass and the age values from the Besan\c{c}on model. To indicate their velocity dispersion, we take into account both the mass-velocity and age-velocity correlations. The AVR is resulted by averaging over stellar mass and the MVR is achieved by averaging over stellar age. Hence, for lens and source stars with the specified mass and age $\mathcal{A}$, we first determine its dispersion velocity according to the MVR, $v_{1}$. This velocity is related to a star with an average age of $\left<\mathcal{A}\right>$, then we shift the AVR by $\Delta=v_{1}-v(\left<\mathcal{A}\right>)$, where $v$ is the size of the stellar velocity dispersion as given by the AVR. Finally, using the shifted equation, which is $v(\mathcal{A}) + \Delta$, we calculate its dispersion velocity. The contributions of different components in this velocity dispersion, i.e., $v_{\rm U},~v_{\rm V},~v_{\rm W}$, are given by the Besan\c{c}on model \citep{Robin2003}. \\


We choose events in our Monte-Carlo simulation based on the observational-based criteria as listed here. (a) The source star (by considering the blending effect) at the baseline should be brighter than $20$ mag. (b) The source stars should be fainter than $14.25$ mag at the peak of magnification in the standard $I$-band filter. (c) The events in the simulation are accepted based on the MOA detection efficiency, as depicted in Figure (2) of \citet{Sumi2013}. We consider the blending parameter as a weight function while calculating average values. Because high-blending events are not discernible in real microlensing observations.

\begin{figure*}
\centering
\includegraphics[width=0.48\textwidth]{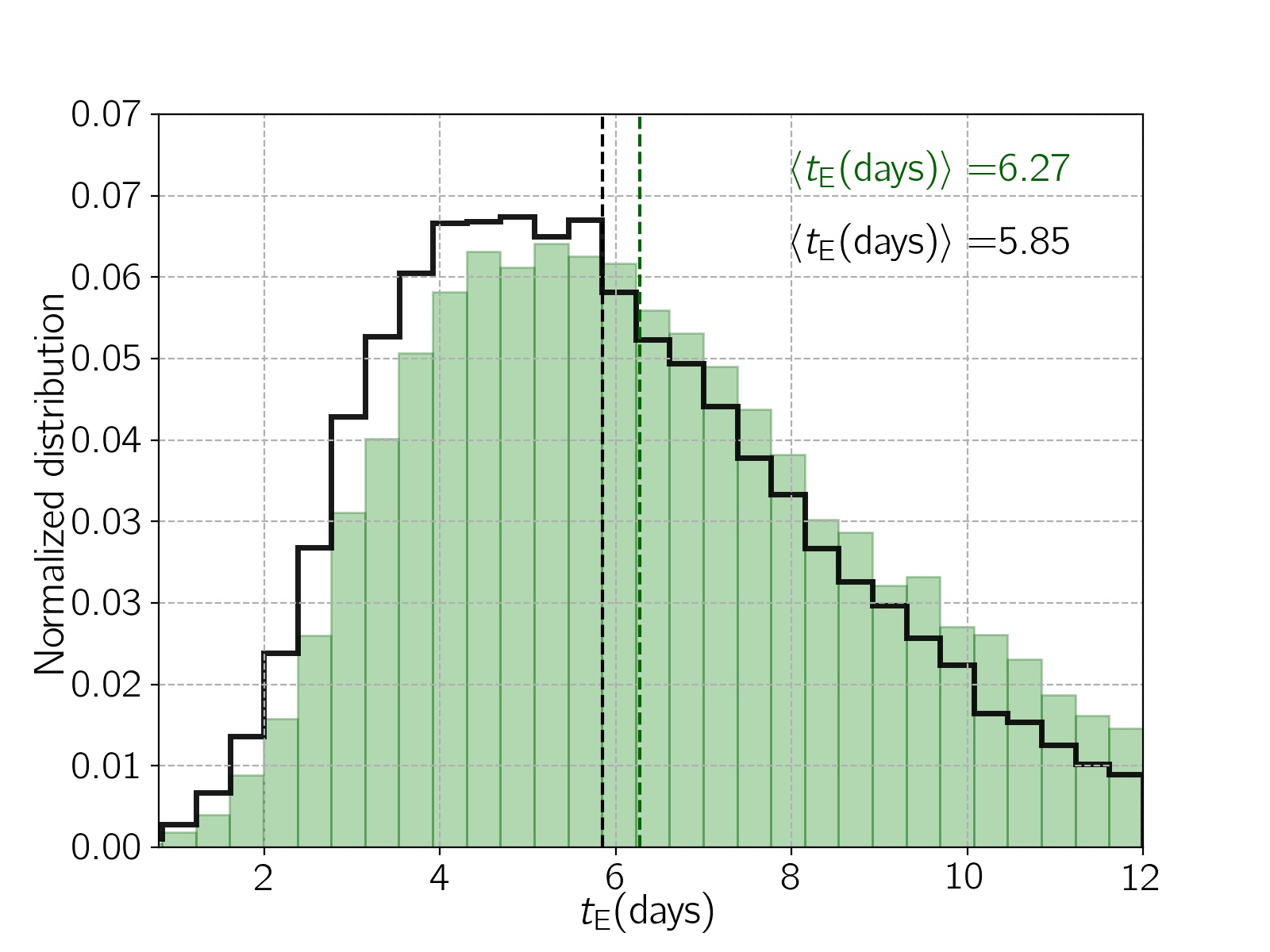}
\includegraphics[width=0.48\textwidth]{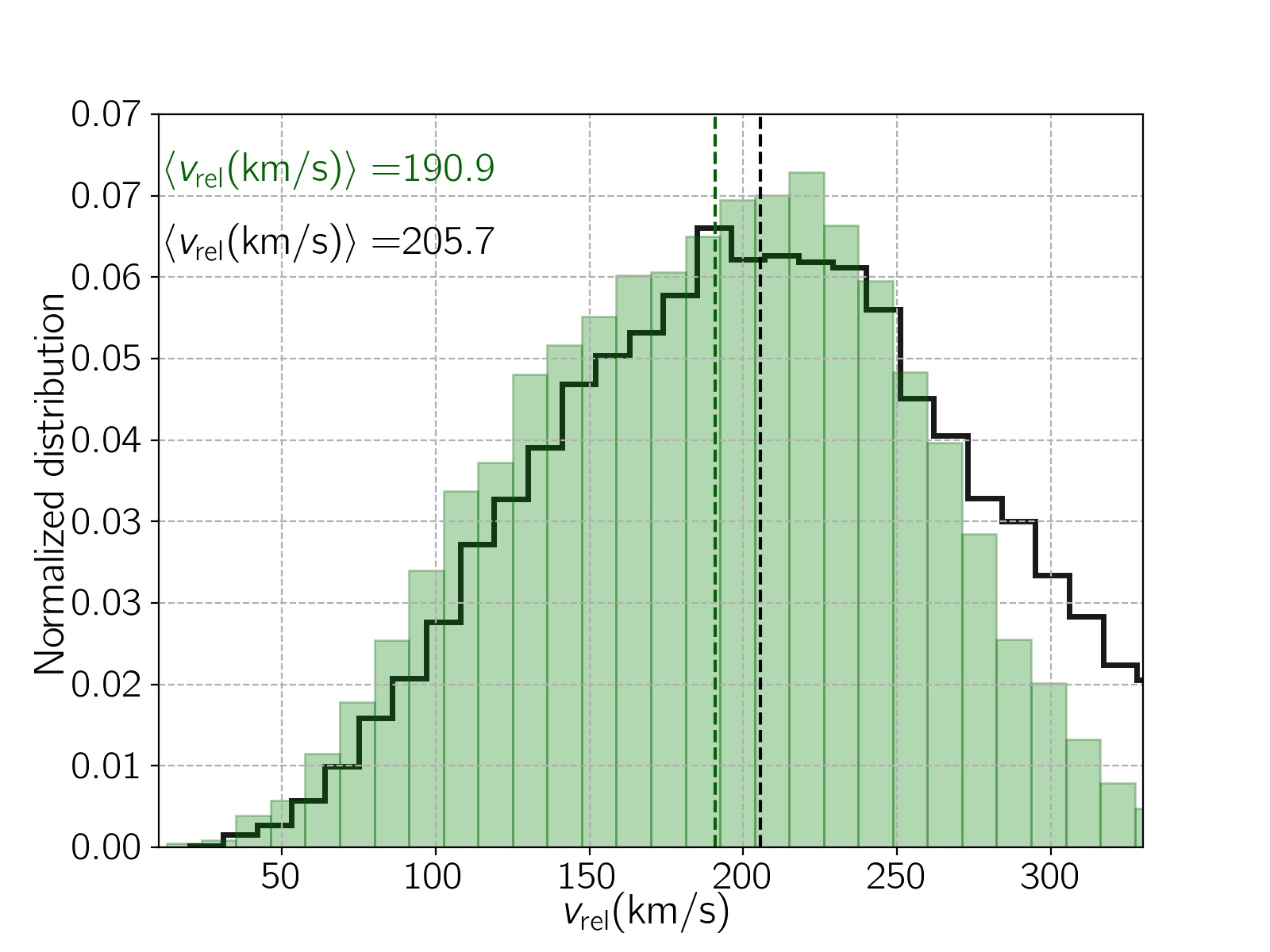}
\caption{The normalized distribution of the Einstein crossing time (left panel) and the lens-source relative velocity (right panel) of simulated microlensing events. In the simulation for each lensing configuration, the lens-source relative velocity and the Einstein crossing time are calculated by considering the MVR (black step distributions) and without applying it (filled light-green ones).}\label{histotE}
\end{figure*}
\begin{deluxetable*}{cccccccccc}
\tablecolumns{10}
\centering
\tablewidth{0.95\textwidth}\tabletypesize\footnotesize\tablecaption{Table shows the effect of the MVR on the lensing parameters.}
\tablehead{\colhead{$t_{\rm E}\rm{(days)}$}& \colhead{} & \colhead{$1.7\pm0.3$} & \colhead{$2.6\pm0.3$} & \colhead{$3.5\pm 0.3$} & \colhead{$4.5\pm 0.3$} & \colhead{$5.5\pm 0.3$} & \colhead{$6.5\pm 0.3$} & \colhead{$7.5\pm 0.3$} & \colhead{$8.5\pm0.3$} } \\
\startdata
\\
$M_{\rm l}(M_{\odot})$& $\rm{(i)}$ &$0.025\pm 0.01$ & $0.025\pm 0.01$ & $0.027\pm 0.01$ & $0.033\pm 0.02$ & $0.039\pm 0.02$ & $0.043\pm 0.02$ & $0.046\pm 0.02$ & $0.047\pm 0.02$ \\
$~$ & $\rm{(ii)}$ & $0.023\pm 0.01$ & $0.025\pm 0.01$ & $0.029\pm 0.01$ & $0.034\pm 0.02$ & $0.040\pm 0.02$ & $0.044\pm 0.02$ & $0.047\pm 0.02$ & $0.048\pm 0.02$\\
\hline
\\
$v_{\rm{rel}}\rm{(km/s)}$ & $\rm{(i)}$ &$234.1\pm 50.2$ & $237.3\pm 49.2$ & $232.4\pm 47.9$ & $217.6\pm 51.0$ & $205.2\pm 53.8$ & $189.9\pm 55.2$ & $174.6\pm 52.5$ & $156.7\pm 48.1$ \\
$~$& $\rm{(ii)}$& $260.8\pm 58.6$ & $261.1\pm 54.9$ & $245.3\pm 55.4$ & $226.3\pm 57.3$ & $213.0\pm 56.8$ & $195.3\pm 56.1$ & $178.1\pm 50.9$ & $161.6\pm 46.6$\\
\hline
\\
$D_{\rm l}\rm{(kpc)}$& $\rm{(i)}$&$4.449\pm 3.40$ & $4.162\pm 3.08$ & $4.211\pm 2.61$ & $4.477\pm 2.44$ & $4.657\pm 2.31$ & $4.890\pm 2.22$ & $5.085\pm 2.12$ & $5.305\pm 2.05$ \\
$ $& $\rm{(ii)}$ & $5.018\pm 3.39$ & $4.679\pm 2.93$ & $4.615\pm 2.58$ & $4.660\pm 2.42$ & $4.787\pm 2.29$ & $4.913\pm 2.19$ & $4.947\pm 2.09$ & $5.128\pm 2.07$ \\
\tablecomments{For timescale values given in the first row, the average values of the lens mass, the lens-source relative velocity, and the lens distance from the observer due to the simulated microlensing events by ignoring the MVR (i), and by applying the MVR (ii) are reported in the next rows, respectively.}
\enddata
\label{tabfin} 
\end{deluxetable*}

In order to show the effect of the MVR on the simulated microlensing events, for each lensing configuration with the specified lens and source stars, we calculate the lens-source relative velocity $v_{\rm{rel}}$ and, as a result, the Einstein crossing time two times: (i) by ignoring the MVR, and (ii) by applying this correlation. In Figure \ref{histotE}, we show the normalized distributions of the Einstein crossing time (left panel), and the lens-source relative velocity (right panel) for the events (i) with filled light-green and (ii) with black step distributions. By employing the MVR, the resulting microlensing events have, on average, larger lens-source relative velocities and shorter timescales. The mean values for each distribution are mentioned in Figure \ref{histotE}.

Accordingly, disregarding the MVR while modeling short-duration microlensing events and doing the Bayesian analysis (when the microlensing degeneracy is not resolvable) causes the lens masses to be underestimated. In Table \ref{tabfin}, for several discrete ranges of the Einstein crossing time, we report the average value of the lens mass, the lens-source relative velocity and the lens distance. Generally, in our simulation for a microlensing event on a given time scale, by taking into account the mass-velocity dispersion relation, the inferred mass from analyzing light curves is larger by the amount of $\delta M_{\rm l}/M_{\rm l}\simeq 2.5$-$5.5\%$. This point emphasizes the importance of the MVR while modeling short-duration microlensing events. \\

\section{Summary and Conclusions}\label{five}

The average value of stellar dispersions increases with stellar age, the known AVR, which is an output of the gravitational interaction of stars with the Galactic giant structures such as spiral arms and massive gas and molecular clouds. Early-type stars have shorter lifetime than the late-type stars. So, the massive and early-type stars are on average younger and have lower dispersions than the late-type stars. However, low-mass and late-type stars, e.g., K and M-type stars, and brown dwarfs have ages as long as the Universe, and according to the AVR these low-mass stars have almost similar dispersion profiles.\\

In a thermalized and collisional N-body system with equipment conditions, e.g., globular clusters in our galaxy, the velocities of the particles obey the MB distribution. In this distribution, low-mass particles move faster than massive stars. However, in our galaxy, stellar collisions with each other and MACHOs are rare, and its relaxation time is of the order of $10~$Gyrs. As a result, we expect that in the galaxy, stellar velocity dispersion depends weakly on mass (a second-order effect). In order to verify this point and find a correlation between the velocity dispersion profile of stars and their mass (even a weak correlation), we used the \gaia~archived data.\\

We took an ensemble of stars which were observed by \gaia,~by considering two restrictions of (a) closer than $150$ pc, and (b) brighter than $21$ mag in $G$-band. We excluded variables and giant stars and took main-sequence stars. The components of their velocity, in the heliocentric frame, projected onto the sky plane and towards the Galactic north and east, and the line of sight direction were measured by the \gaia~telescope. We first transformed these velocity dispersions from the heliocentric frame to the LSR frame. Because the size of our sample is small (closer than $150$ pc) and the observer has the same co-moving velocity, the \gaia~telescope has measured the velocity dispersion of these stars.


\noindent By knowing the stellar mass and dispersion velocity of stars, we made several subsamples of these stars with discrete mass ranges (and the same number of entrances), plotted the density profiles of their velocities and fitted MB distributions to them. We noticed that there was a correlation between the average mass $\left<M\right>$ of stars in each subsample and the scale parameter of $a$ in the best-fitted MB distribution. According to the properties of the MB distribution, the scale parameter increases by enhancing $1/ \sqrt{\left<M \right>}$. We fitted two linear relations to the data as given by Equation \ref{corre} and plotted in Figure \ref{fitted}. Since our galaxy is not a completely thermalized system, we expect the relation between the stellar velocity dispersion and $1/ \sqrt{\left<M \right>}$ is not completely linear, especially for low-mass subsamples. Hence, we fitted a power-law equation to the data with $\left<M\right> \lesssim 0.98 M_{\odot}$ as given by Equation \ref{corre2}.

As a result of this correlation, which is nominated as the MVR, brown dwarfs have a wider dispersion profile with the extrapolated scale parameter $a \simeq 54~\rm{km/s}$, than main-sequence stars. The MVR confirms the known AVR (while comparing the velocity of early and late-type stars, and when $\left<M\right> \gtrsim 0.98 M_{\odot}$). Furthermore, it gives us another weak correlation for velocity dispersions of low-mass stars and brown dwarfs with their mass. We emphasize that the MVR is only at $2$-$3$ sigma level and the proposed extrapolation to the BD regime is hypothetical.\\

This correlation affects interpreting short-duration and degenerate microlensing events. To show this point and evaluate its effect, we performed a Monte-Carlo simulation. We made an ensemble of synthetic microlensing events toward the Galactic bulge due to brown dwarfs. We found that taking into account the MVR in modeling these microlensing events using the Bayesian analysis results in correction to the lens mass by $\sim 2.5$-$5.5\%$. \\

\section*{Acknowledgments}
This work has made use of data from the European Space Agency (ESA) mission {\it Gaia} (\url{https://www.cosmos.esa.int/gaia}), processed by the {\it Gaia} Data Processing and Analysis Consortium (DPAC, \url{https://www.cosmos.esa.int/web/gaia/dpac/consortium}). Funding for the DPAC has been provided by national institutions, in particular the institutions participating in the {\it Gaia} Multilateral Agreement. We acknowledge V.~Bozza and R.~Poleski for reading the paper and commenting on it. We also thank H. Fatheddin for his help to prepare some part of data and the anonymous Referee for his/her careful comments which improved the quality of the paper. S.~Sajadian thanks the Department of Physics, Chungbuk National University and especially C.~Han for hospitality.\\

\bibliography{references}{}
\bibliographystyle{aasjournal}

\end{document}